# Momentum spectroscopy for multiple ionization of cold rubidium in the elliptically polarized laser field*

Junyang Yuan(袁俊阳)[1,2,3], Yixuan Ma(马祎璇)[1,2,3], Renyuan Li(李任远)[1,2], Huanyu Ma(马欢玉)[1,2,3], Yizhu Zhang(张逸竹)[1,4], Difa Ye(叶地发)[5], Zhenjie Shen(沈镇捷)[1], Tianmin Yan(阎天民)[1**], Xincheng Wang(王新成)[3**], Matthias Weidemüller[6,7,8], Yuhai Jiang(江玉海)[1,2,3,7**]

[1] Shanghai Advanced Research Institute, Chinese Academy of Sciences, Shanghai 201210

[2] University of Chinese Academy of Sciences, Beijing 100049

[3] School of Physical Science and Technology, ShanghaiTech University, Shanghai 201210

[4] Center for Terahertz Waves and College of Precision Instrument and Optoelectronics Engineering, Key Laboratory of Education, Tianjin University, Tianjin 300072

[5] Laboratory of Computational Physics, Institute of Applied Physics and Computational Mathematics, Beijing 100088

[6] Hefei National Laboratory for Physical Sciences at the Microscale and Shanghai Branch, University of Science and Technology of China, 201315 Shanghai, China

[7] CAS Center for Excellence and Synergetic Innovation Center in Quantum Information and Quantum Physics, University of Science and Technology of China, 201315 Shanghai, China

[8] Physikalisches Institut, Universität Heidelberg, Im Neuenheimer Feld 226, 69120 Heidelberg, Germany

*Supported by the National Natural Science Foundation of China (NSFC) (11827806, 11874368, 61675213, 11822401, 11674034). We acknowledge the support from Shanghai-XFEL beamline project (SBP) and Shanghai High repetition rate XFEL and Extreme light facility (SHINE).

**Email: jiangyh@sari.ac.cn; wangxch1@shanghaitech.edu.cn; yantm@sari.ac.cn

%Tel. 021-20350819 13651699807 18321614135 17702108138

**Abstract** Employing recent developed magneto-optical trap recoil ion momentum spectroscopy (MOTRIMS) combining cold atom, strong laser pulse, and ultrafast technologies, we study momentum distributions of the multiply ionized cold rubidium (Rb) induced by the elliptically polarized laser pulses (35 fs, $1.3 \times 10^{15}$ W/cm$^2$). The complete vector momenta of Rb$^{n+}$ ions up to charge state $n = 4$ are recorded with extremely high resolution (0.12 a.u. for Rb$^+$). Variations of characteristic multi-bands displayed in momentum distributions, as the ellipticity varies from the linear to circular polarization, are interpreted qualitatively with the classical over-barrier ionization model. Present momentum spectroscopy of cold heavy alkali atoms presents novel strong-field phenomena beyond the noble gases.

**PACS:** 32.80.Fb, 34.50.Fa, 32.80.Qk, 42.50.Hz

---

The interaction of strong laser fields with atoms and molecules has attracted a large number of exciting phenomena in forefront strong-field physics, among which are above-threshold ionization (ATI),[1-4] sequential double ionization (SDI) and nonsequential double ionization (NSDI),[5, 6] high-order harmonic generation[7, 8] and attosecond pulses creation.[9-12] To date, most theoretical and experimental studies above are devoted to linearly polarized (LP) fields. Recently, a growing interest is focused on the study of strong-field ionization induced by elliptically polarized (EP) fields.[13-15] As EP fields has the ability to uncover ionization information that is unreachable with LP fields. For example, by measuring the electron momentum from SDI in the EP fields, one can obtain information for the ionization fields and the release times of the electrons,[16-18] which cannot be directly or easily obtained with LP fields. However, this can be straightforwardly obtained from the end-of-pulse recoil-ion momentum distributions (RIMDs) obtained under EP fields, even without the electron-ion coincidence detection.

Experimental techniques such as cold target recoil ion momentum spectroscopy (COLTRIMS)[19] and velocity map imaging (VMI)[20] have been used to measure

RIMDs. In these techniques, an essential step is to precool the atom target for the acquisition of RIMDs with high resolution. Most experiments focus on targets of rare gas atoms or molecular gases, because these gases can be cooled efficiently by supersonic expansion of gas jet targets. Alkali atoms, on the other hand, have rarely been studied under the context of strong field processes, though the alkali atoms are featured for their low ionization energies and the easy manipulation of Rydberg states according to their solid phase in the room temperature. The heating sublimation to gas phase will lead to poor momentum resolution for recoil-ions. In fact, the preparison of cold alkali atoms with magneto-optical-trap (MOT) techniques is a routine in laboratories of cold atom physics although involved techniques are extremely high. Combining the MOT technique with a COLTRIMS, an integration of ultra-cold atom, strong laser pulse, and ultrafast technology, we set up a MOTRIMS platform,[21] availing the study of ultrafast processes within alkali or alkaline-earth atoms that otherwise cannot be cooled by conventional supersonic expansion. In atomic, molecular, and optical (AMO) physics, MOTRIMS allowed to study various aspects of the electron capture process: examples are inner-shell and outer-shell electron captures,[22, 23] photoassociation in cold atoms,[24] ion-atom/molecule and photon atom/molecule collisions.[25] Beyond AMO physics, MOTRIMS is also used for precision measurements in nuclear physics.[26, 27] To date, there is no MOTRIMS apparatus combined strong and ultrafast laser fields reported in the literature. Thus, the combination of these three advanced techniques–COLTRIMS, MOT, and the femtosecond pulse–represents an acutely powerful experimental tool for the study of momentum spectroscopy for heavy alkali atoms in the strong-field physic, being novel investigations in the multi-dimensions beyond the noble gases.[28]

In this Letter, we study the strong field multiple ionization of Rb targets in the EP fields with our MOTRIMS setup. The neutral Rb atoms are ionized up to $Rb^{4+}$ and the RIMDs are reconstructed with extremely high resolution (0.12 a.u. for $Rb^+$). The momentum distributions exhibit rich multi-band structures as the ellipticity varies from the LP to close-to circularly polarized (CP) fields. With the help of the classical over-barrier ionization model, we analyze experimental observations and identify underlying mechanisms.

The experimental setup is based on a double chamber system, with a Rb atom source prepared in a glass cell and then delivered into the target region of the science chamber. In the glass cell, Rb atoms are pre-cooled within a typical two-dimensional magneto-optical trap (2D MOT) configuration, and then pushed by a laser beam into the target region, where it can be further cooled and trapped with a standard three-dimensional magneto-optical trap (3D MOT) configuration. Either the 3D MOT target, molasses or the 2D MOT target with various densities is selected within the target region. Detailed information about the experimental setup and target preparation is presented in Ref. [21]. The density of the molasses target used in this experiment is approximate $10^8$ atoms/cm$^3$, and the background vacuum in the science chamber is lower than $2\times10^{-10}$ mbar. The intense femtosecond laser pulses (~35 fs, 800 nm) are generated from a mode-locked Ti: sapphire laser system with a repetition rate of 1 kHz. A combination of a $\lambda/2$ plate and an alpha-BBO Glan-Taylor laser polarizer are used to adjust the incident laser intensity, and the peak intensity $I$ is calibrated by measuring the "donut"-shape RIMDs of doubly charged Rb$^{2+}$ with CP fields below the over-barrier ionization.[29] The uncertainty of $I$ is estimated to be 20%. A zero-order quarter-wave plate is positioned at the entrance of the target chamber to polarize the laser with ellipticity $\varepsilon$, then the polarzied laser beam is focused by a concave mirror of 75 mm focal length onto the cold Rb molasses target. During the experiment, varing the ellipticity at a constant intensity $1.3 \times 10^{15}$ W/cm$^2$, the Keldysh parameter $\gamma$ changes from 0.42 to 0.82 for all involved charge states. After ionization, the ions are accelerated by a homogeneous electric field (~ 1 V/cm) along the time-of-flight (TOF) axis toward a time- and position-sensitive microchannel plate (MCP) detectors. The channel plate detector followed by a multihit anode allows extracting the information, including the TOFs and the positions of the arrival ions, by which the full vector momenta of these ions can be reconstructed.

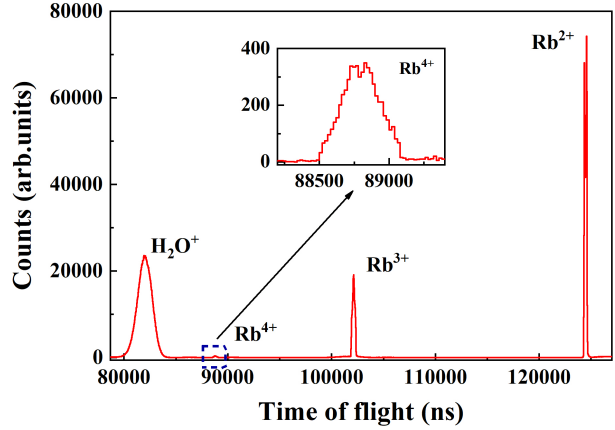

Fig. 1. The time-of-flight (TOF) spectrum of Rb ions. The peaks from right to left show the signals of $Rb^{2+}$, $Rb^{3+}$, $Rb^{4+}$ and $H_2O^+$ ions in intense close-to CP fields at a peak intensity of $1.3 \times 10^{15}$ W/cm$^2$ (background pressure lower than $2 \times 10^{-10}$ mbar), repectively.

A typical TOF spectrum for different charge states of $Rb^{n+}$ is shown in Fig. 1. The recoil ions are selected in a time-of-flight (TOF) mass spectrometer according to their mass-to-charge ratio. Here, the $Rb^+$ ions are not collected during data acquisition, since the extremely high ionization rate results in the saturation of detectors and very slow data processing. The FWHMs (full width half maximum) for the peaks of $Rb^{2+}$, $Rb^{3+}$, $Rb^{4+}$ are about 286 ns, 319 ns, 330 ns, respectively, which are at least one order narrower than $H_2O^+$ peak. Note that one has to keep physical broadening of these peaks in mind. Here, taking $Rb^+$, a resolving power $m/\Delta m$ of 3000 was optimized in the present experimental settings.

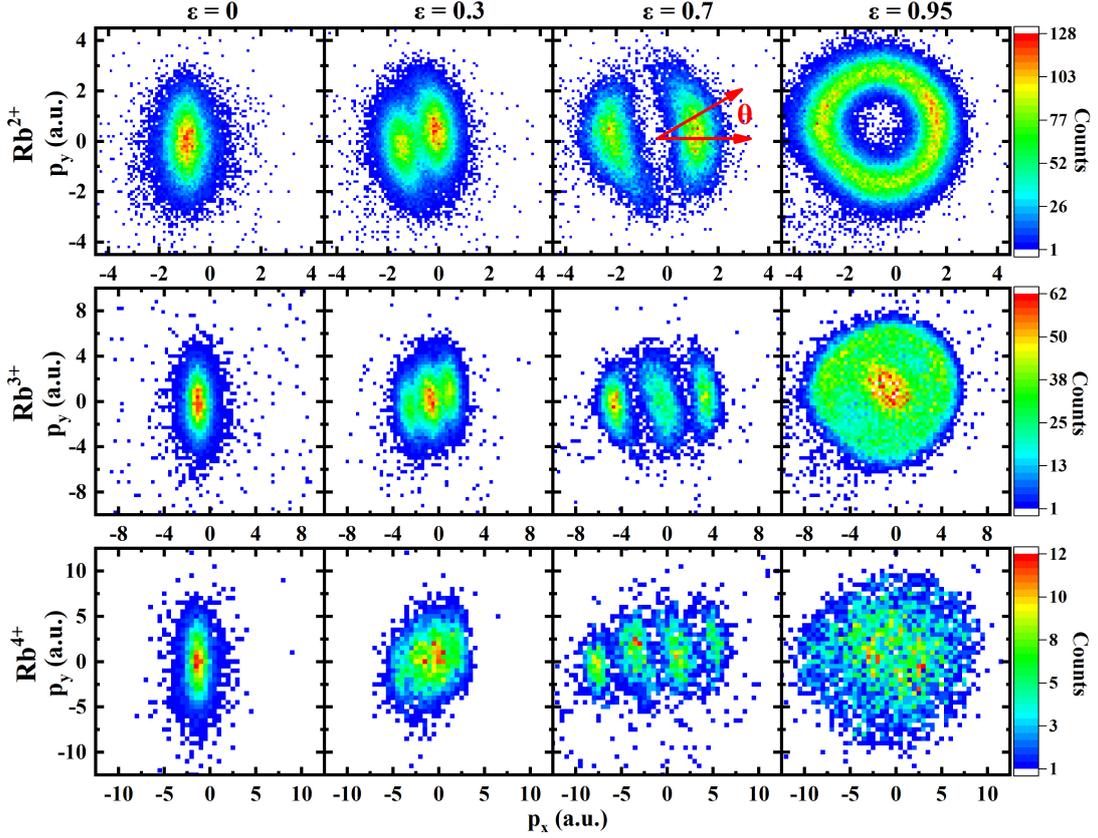

Fig. 2 (color online). Measured RIMDs in the polarization plane for different ellipticities $\varepsilon$ (columns) from LP to close-to CP fields for double, triple, and quadruple ionizations (rows) of neutral Rb atoms. The major polarization and the minor polarization axis are along the $y$ and $x$ axises, respectively. The offset angle $\theta$, indicated in the panel for $\varepsilon = 0.7$, is defined as the angle between the maximum of the RIMDs and the minor polarization axis of the laser electric field. The peak intensity of the laser pulse is about $1.3 \times 10^{15}$ W/cm$^2$.

The $\varepsilon$-dependent RIMDs for Rb$^{2+}$, Rb$^{3+}$ and Rb$^{4+}$ in the polarization $x$-$y$ plane, where $y$-axis and $x$-axis are defined along the major and minor polarization axis, respectively, are shown in Fig. 2. For $\varepsilon = 0$ (the linear polarization along the y-axis), the RIMDs display two-dimensional Gaussian-like distributions with the maxima at zero position. As expected, the distribution is expanded in the polarization direction (y-axis). For Rb$^{2+}$, Rb$^{3+}$ and Rb$^{4+}$ ions, the ratios of the RIMDs along the major axis to the minor axis are 2.2, 3.0 and 3.5, respectively. As ellipticities increase, the RIMDs split into multi-band structures along the minor axis ($x$ axis) of the polarization ellipse and the gap between each peaked band increases with the ellipticity. At $\varepsilon = 0.7$, two-, three- and four-band structures are presented for $Rb^{2+}$, $Rb^{3+}$ and $Rb^{4+}$, respectively. The distributions are more likely to be clustered around the minor axis. This is due to the fact that the amplitudes of the electric field in the *y* direction are stronger than those in the *x* direction, and thus the electrons prefer emission along the *y* axis. The electron that emits along the *y* axis at the times of *y* maximum of the electron field achieves a final momentum with a large *x* component.[30, 31] In the case of close-to CP fields, the RIMDs of $Rb^{2+}$, $Rb^{3+}$, and $Rb^{4+}$ show a circularly symmetric structure, two concentric rings, and a connected one part structure, respectively. Moreover, these distributions are relatively isotropic.

For $Rb^{2+}$ at $\varepsilon = 0.7$, the two peaks are shifted by an offset angle $\theta$ marked in Fig. 2, that is, the angle between the maximum of the RIMDs and the minor axis of the polarization ellipse. This angles around 45°-55°decreases with increasing ellipticities. Recently, much attention has been concentrated on exploring the physical mechanism of this angle shift. It is believed that the $\theta$ angle is wholly or partially caused by the long-range Coulomb interaction between the outgoing electron and the ionic core.[13] In addition, some explanations attribute the angle to the nonzero initial momentum at the tunnel exit in the polarization plane, and the angle shift relies heavily on the initial conditions of the electron right after the tunneling in the model, such as its momentum, position and/or a finite tunneling time.[32, 33] This angle shift has been widely analyzed in attoclock measurements, which allows resolving photoelectron dynamics with attosecond precision with available femtosecond laser pulses.[13] However, all these explanations are dependent on the model employed, and it is still controversial to draw the final conclusion.

In order to better understand the physical mechanism of these shape characteristics, a quantitative explanation of our experimental results in terms of over-barrier ionization in close-to CP fields is presented as followings. First, we will give a brief description of the classical over-barrier ionization. Taking an electron residing in one-dimensional potential along the field direction for instance,[34] the quasi-static electric field of increasing strength will further suppress the Coulomb barrier until the electron is no more bound when reaching the critical field strenth $I_p^2/4Z$. Here $I_p$ is the ionization energy. Assuming that each electron is ionized with zero initial momentum and the ion-core Coulomb attraction can be neglected after ionization, the momentum of each electron at the end of the pulse is given by $p = I_p^2/4Z\omega$. The direction of the final momentum is perpendicular to **E**. At this time, the corresponding field intensity can be estimated as $I_{OBI} = I_p^4/16Z^2$. According to the explanation above, the over-barrier intensity $I_{OBI}$ of $Rb^+$, $Rb^{2+}$, $Rb^{3+}$ and $Rb^{4+}$ ions can then be estimated as: $1.8 \times 10^{11}$ W/cm², $5.6 \times 10^{14}$ W/cm², $1.0 \times 10^{15}$ W/cm², $1.9 \times 10^{15}$ W/cm², respectively. In this case the $Rb^+$, $Rb^{2+}$ and $Rb^{3+}$ ions are generated well above the over-barrier intensity while $Rb^{4+}$ ions are generated around the over-barrier region for the laser peak intensity used in our experiment. It should be noted that in the over-barrier region, the RIMDs no longer depend on the laser intensity. Indeed, we have further examined spectra of a wide range of higher laser intensities and almost identical results are obtained.

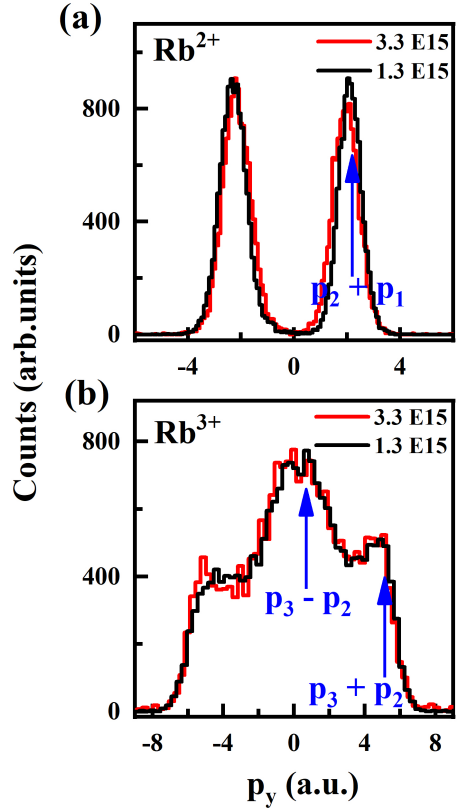

Fig. 3. RIMDs of $Rb^{2+}$ [panel (a)] and $Rb^{3+}$ [panel (b)] ions projected onto the y axis (along the TOF direction) for close-to CP fields at peak intensity of $1.3 \times 10^{15}$ W/cm² (black) and $3.3 \times 10^{15}$

W/cm$^2$ (red). The vertical arrows in Fig. 3(a) and (b) indicates the locations expected for the over-barrier ionizations, as discussed in the text.

To be more intuitive, Fig. 3 shows the $p_y$ spectra (in the TOF direction where the best resolution is achieved) of Rb$^{2+}$ and Rb$^{3+}$ ions at peak intensities of $1.3 \times 10^{15}$ and $3.3 \times 10^{15}$ W/cm$^2$. It is clear that the shape of the spectra is independent with laser intensities, just as expected in the over-barrier region. Moreover, the Rb$^{2+}$ ions show a double-peak structure and the Rb$^{3+}$ ions show a four-peak structure. According to classical calculations for the noble gas system, the momentum spectrum exhibit double-, four- and eight-peak structures for singly, doubly and triply charged ions, respectively,[18, 35] due to the different combinations of emission directions for the emitted electrons. The observed momentum distribution of Rb$^{2+}$ (Rb$^{3+}$) is equivalent to that of the singly (doubly) charged ion as predicted by classical model. It is because that for Rb the first and second ionization potentials are 2.6 eV (excited state) and 27.3 eV, respectively, corresponding to over-barrier fields of 0.0023 a.u. and 0.1259 a.u., the corresponding values of $p_1$ and $p_2$ are given by about 0.04 a.u. and 2.21 a.u.. If the two electrons were released only as the instant the laser field reached these two values, not only the magnitudes of $p_1$ and $p_2$ would be unique, but also the angle between them as well. Note that a much more likely scenario is that the electrons were released over a range of times when the field is near the over-barrier value.[35] If this range of times is larger than the optical cycle, the corresponding range of angles will be more than $2\pi$. In this case one obtains recoil momentum which range between $p_2 + p_1 = 2.2492$ a.u. (when the two electrons emission into the same direction) and $p_2 - p_1 = 2.1690$ a.u. (when into opposite directions). Averaging over all possible angles between $p_1$ and $p_2$, when projected onto $p_y$, the RIMDs will lie inside and peaks near 2.25 a.u. and 2.17 a.u.. However, since the difference (about 0.0802 a.u.) between $p_2 + p_1$ and $p_2 - p_1$ is even smaller than the instrumental resolution (~0.12 a.u. for Rb$^+$), the double-peak structure is indistinguishable for the peaks around ±2.20 a.u. as shown in Fig. 3(a). From the discussion above, the following conclusions are drawn: the momentum of the valence electron is so small that its effects can be ignored when analyzing the ionization mechanism of higher charge states.

For $Rb^{3+}$, the third ionization potentials of Rb is 39.2 eV, corresponding to momentum of the third electron $p_3$ = 3.0365 a.u.. When the second and the third electrons exit in the same direction, the total momentum of Rb ion core $p_{ion}$ is the sum of momenta of the second and third electrons, for the $Rb^{3+}$ ion core, $p_{ions} = p_3 + p_2$ = 5.2456 a.u.. When the two electrons emit in the opposite direction, the total momentum of the ion core, however, is $p_{ions} = p_3 - p_2$ = 0.8274 a.u.. As shown in Fig. 3(b), the $Rb^{3+}$ spectra show a four-peak structure near ±0.8 a.u. (two inner peaks) and near ±5.2 a.u. (two outer peaks).

In summary, employing recent developed MOTRIMS combining cold atom, strong laser pulse, and ultrafast technologies, multiple ionizations up to quadruple ionization of neutral Rb target by EP fields were investigated. The ion momentum spectrum exhibit characteristic multi-band structures as the ellipticity varies from the LP to CP fields. Theoretical analysis shows that the momentum of the valence electron is so small that its effects can be ignored when analyzing the physical mechanism of higher charge states, this can be verified from the RIMDs of $Rb^{2+}$ by close-to CP fields. Accordingly, the spectra of $Rb^{3+}$ can be interpreted quantitatively in terms of two successive classical over-barrier ionization model. The quantitative agreement between the classical model and our experiment results provides strong support to the classical treatment of the multielectron processes induced by strong laser fields, which is currently indispensable because the nonperturbative quantum treatments of the complex effect are not feasible.